\documentclass[twocolumn,showpacs,showkeys,superscriptaddress]{revtex4}

\usepackage{bm}
\usepackage{color}
\usepackage{amsmath}
\usepackage{natbib}

\begin{document}

\title{Generating vorticity and magnetic fields in plasmas in general relativity: spacetime curvature drive}

\author{Felipe A. Asenjo}
\email{faz@physics.utexas.edu}
\affiliation{Institute for
Fusion Studies, The University of Texas at Austin, Texas 78712,
USA.}

\author{Swadesh M. Mahajan}
\email{mahajan@mail.utexas.edu}
\affiliation{Institute for Fusion Studies, The University of Texas at Austin, Texas 78712, USA.}
\author{Asghar Qadir}
\affiliation{Centre for Advanced Mathematics \& Physics, National University of Sciences \& Technology, H12, Islamabad 4400, Pakistan.}

\date{\today}

\begin{abstract}
Using the generally covariant magnetofluid formalism for a hot
plasma, a new spacetime curvature  driven mechanism for
generating seed vorticity/magnetic field is presented. The
``battery'' owes its origin to the interaction between gravity
and the inhomogeneous plasma thermodynamics. The general
relativistic drive is evaluated  for two simple cases: seed formation in
a simplified model of a hot plasma accreting in stable orbits around a
Schwarzschild black hole, and for particles in free fall near the horizon.
Some astrophysical applications
are suggested.
\end{abstract}

\pacs{04.70.Bw, 52.27.Ny, 52.35.We, 95.30.Qd}


\maketitle


\section{Introduction}

Just as the motion of a charged fluid in space-time generates a
magnetic field, it stands to reason that if spacetime were
distorted in the region occupied by a charged fluid, a magnetic
field would emerge. In a special relativistic context, it was recently demonstrated \cite{mah1,mah12}
that a generalized vorticity (GV)
\begin{equation}
\hat{\bf{B}} = {\bf B} + \frac{m}{q} \nabla\times (f \gamma\bm{v}),
\label{GV}
\end{equation}
consisting of magnetic and kinetic-thermal parts, may be
generated, $\it{ab initio}$, in an ideal perfect fluid with
inhomogeneous entropy. In \eqref{GV}, $\bm{v}$ is the velocity,
$m$ ($q$) are, respectively, the mass (charge) of the fluid
particle, $ \gamma=(1-v^2/c^2)^{-1/2}$ is the Lorentz factor,
 $v^2=\bm{v}\cdot\bm{v}$, $c$ is the speed of light and $f$ is
the relativistic thermal factor related to the fluid density
enthalpy $h=n mc^2 f$, with $n$ as the fluid density. For a
relativistic Maxwell distribution, $f\equiv f(x)=K_3(x)/K_2(x)$
\cite{L-L,dzhav}, where $K_j$ are the modified Bessel functions of
order $j$, and $x=mc^2/k_B T$ is the inverse normalized
temperature with the Boltzmann constant $k_B$.  It is to be emphasized that the GV generation in special relativity proposed in Refs.~\cite{mah1,mah12} is entirely
due to a distortion of {\it space} (as distinct from spacetime)
caused by the special relativistic $\gamma$-factor. It is well known
that in the non-relativistic dynamics of an ideal fluid, a
topological constraint would forbid the emergence of GV from a zero
initial value. Of course, motion in one frame need not be motion in
another and so the distortion is frame-dependent. If the
astrophysical choice of ``rest-frame" is clear the frame-dependence
need not worry one.

In this paper we explore the possible role of general relativity in the generation of magnetic fields in plasmas. The aim is to find a generalization for the magnetic field generation in special relativity discussed above.

A somewhat different interpretation of this special relativistic effect
will be helpful in casting  light on the extension to curved spacetime.
The Poincar\`{e} group, $SO(1,3)\bigotimes_s \Re^4$, where
$\bigotimes_s$ is the semi-direct product, guarantees the
conservation of energy and momentum via spacetime translational
invariance ($\Re^4$). Since $SO(1,3)\cong SO(3)\bigotimes SO(3)$,
the first, $SO_L(3)$, can yield angular momentum conservation,
the second, $SO_S(3)$, will give the conservation of Dirac's
{\it spin angular momentum conservation}. The former
corresponds to spatial rotational invariance and the second to
proper Lorentz invariance. Of course, what is conserved is the total angular
momentum ({\bf J} = {\bf L} + {\bf S}), and it is {\it
this} that provides the ``seed" for magnetic field generation
\cite{mah1}.

Notice that, though the rotation can be undone over the entire
spacelike hypersurface in a homogeneous spacetime (hence the
effect is frame dependent for a homogeneous stress-energy
tensor), it will persist in an inhomogeneous system; undoing the
rotation at one place will simply push the twist elsewhere. Even then,
there would have been no ``seed" creation if there were no charge to
induce non-homogeneity in the spacetime;
the plasma is needed to provide the effect. It is also worth pointing out that the
distortion is purely in the spacelike section (as the spacetime
remains flat) and could be locally undone by a change of frame.
However, it cannot be globally undone because of the
inhomogeneity.

In general relativity, however, the curvature of spacetime will provide
an effective motion at one point relative to a ``rest-frame" at
another. More precisely, we can take the local rest-frame at
one point, as given by the tangent space using Riemann normal
coordinates \cite{jq}, and compare it with the local
rest-frame at another point. There will be a definable local
Lorentz factor there, giving the above special relativistic
effect produced by gravity.  The frame
chosen is a special Fermi-Walker frame, which gives the geodesics
{\it as if} they were straight lines bent due to an (appropriately
modified) force of gravity. The GR effects open up the exciting
possibility of spontaneous generation of magnetic fields near
gravitating sources.

In the present calculation we do not consider the back-reaction of the plasma on spacetime.
A complete self-consistent analysis would require the inclusion of
the plasma contribution to the stress-energy tensor that drives the Einstein equations. We expect that
the simpler model, invoked here, will be enough to extract the qualitative features of magnetic field generation in the vicinity of strongly gravitating  bodies.

We begin, in Sec.~\ref{dynamics}, by
writing down the general relativistic plasma equations in a unified form; the word ``unified'' is used in the spirit
of Refs.~\cite{mah1,mah12}. In Sec.~\ref{vorticalGV} we will derive an equation for the generalized vorticity (GV)
that includes the general relativistic (GR) drives for the seed magnetic field. The extension of the special relativistic vortical
dynamics derived in \cite{mah2} (and investigated for vorticity generation in Refs.~\cite{mah1,mah12}) to GR will be accomplished via a $3+1$ decomposition of the plasma equations onto timelike and spacelike hypersurfaces.
 In Sec.~\ref{aplicacion} we estimate the ``value''  of the generated vorticity (magnetic field) seed, and finally in Sec.~\ref{discusion} we provide a perspective  for the results.

\section{Plasma Dynamics}
\label{dynamics}

The dynamics of an ideal plasma (charged fluid) is obtained using the conservation equation for the energy-momentum tensor   $T^{\mu\nu}$ (using the usual symbol $;$  for covariant derivatives)
\begin{equation}
{T^{\mu\nu}}_{;\nu}=q n F^{\mu\nu}U_\nu\, ,
\label{GRenmomen}
\end{equation}
where $F_{\mu\nu}$ is the electromagnetic field tensor and $U^{\mu}$ is the normalized plasma four-velocity ($U^\mu U_\mu=-1$).  Here, we use $c=1$. The charge $q$ and the mass $m$ of the fluid particles are invariants. The energy-momentum tensor for an ideal plasma
\begin{equation}
 T^{\mu\nu}=hU^\mu U^\nu+p g^{\mu\nu}\, ,
\label{GRenmomentensor}
\end{equation}
involves two thermodynamic scalars, the enthalpy density $h$ and the pressure $p$.

The equation of motion \eqref{GRenmomen} could be written in terms of unified fields \cite{mah2} (see also \cite{Bek}). In addition to facilitating calculations, this approach will help us identify GV in general relativity. Invoking the the continuity equation $\left(nU^\mu\right)_{;\mu}=0$, and  introducing the auxiliary thermodynamic function $f=h/mn$, Eq.~\eqref{GRenmomen} is written as
\begin{equation}
 mn U^\nu\left(fU^\mu\right)_{;\nu}=qnF^{\mu\nu}U_\nu-p{,_\nu} g^{\mu\nu}\, .
 \label{Intermediate}
 \end{equation}
Following \cite{mah2}, we define the fully antisymmetric fluid tensor $S^{\mu\nu}=(fU^{\nu})^{;\mu}-(fU^{\mu})^{;\nu}$, and manipulate \eqref{Intermediate} to derive
\begin{equation}
q\ U_{\nu} M^{\mu\nu}=T\sigma^{,\mu}\, ,
\label{eqmo}
\end{equation}
which is the unified covariant equation of motion in terms of the magnetofluid field $M^{\mu\nu}=F^{\mu\nu}+(m/q)S^{\mu\nu}$. All kinematic and thermal (through $f$) aspects of the fluid are now represented by  $S^{\mu\nu}$. The function $\sigma$ is the entropy density of the fluid, and it is related to pressure through
\begin{equation}
 \sigma^{,\mu}=\frac{p^{,\mu}-mn f^{,\mu}}{nT}\, ,
\end{equation}
where $T$ is the temperature. The antisymmetry of $M_{\mu\nu}$ guarantees that the fluid is isentropic $U_{\mu}\sigma^{,\mu}=0$.

Inclusion of the Maxwell equations
\begin{equation}
 {F^{\mu\nu}}_{;\nu}=4\pi q n U^\mu\, ,
 \label{Maxwellcurved}
\end{equation}
completes the system description.

\section{Generation of vorticity and magnetic fields}
\label{vorticalGV}

The main goal of this work is to work out the effects of spacetime curvature (interacting with inhomogeneous entropy) on the special relativistic results \cite{mah1,mah12} on magnetic field generation. Following the standard plasma procedure of Refs.~\cite{tajima,tarkenton,thorne}, we will invoke the spacetime decomposition. The  $3+1$  formalism  allows us to obtain a set of equations that is similar to those found in special relativity, and helps our intuition. It is, perhaps, the main reason  for the popularity of the $3+1$ framework in formulating and solving plasma physics problems in curved spacetime(see for example Refs.~\cite{pla1,pla3,pla4,pla5,
pla2,pla6,pla7,pla8,tajima,tarkenton}).

In the metric tensor in the canonical formalism \cite{misner},
\begin{equation}
ds^2=-\alpha^2dt^2+2\beta_idx^idt+\gamma_{ij}dx^idx^j\, ,~~~(i,j=1,2,3)
\label{eqcm}
\end{equation}
$\alpha$ is the lapse function, $\beta_i$ the shift vector and
$\gamma_{ij}$ is the 3-metric of the spacelike hypersurfaces of metric $g_{\mu\nu}$. Since the square of the lapse
function is the metric component $-g_{00}$, it essentially
corresponds to the gravitational potential. More precisely,
it has been shown that in a particular preferred
frame, called the pseudo-Newtonian frame \cite{aq,qz}
(essentially a special choice of a Fermi-Walker frame), the
gravitational potential comes out to be $\mbox{ln}\sqrt{\alpha}$.  The shift vector
corresponds to the momentum. Of course, in the rest-frame (which can be obtained by an appropriate choice of gauge) the momentum is zero. Assuming that we can still obtain a global
coordinate basis (which will not be possible for the Kerr
metric, for example), we use the rest-frame so as to eliminate
the shift vector. Though the more general discussion is
physically very relevant, in this paper we will limit our
investigations to spacetimes in which the shift vector can
consistently be set to zero; a more complete analysis will betaken up
in a future paper. Note that we could have chosen a frame of
reference (gravitational gauge) to make the lapse function unity
\cite{misner} but this would ``throw the baby out with the
bath-water" as it would not display the gravitational
potential for us to see the physics of its effect on the plasma.
We would then be in the freely falling rest-frame and locally
see Minkowski space around us. This is the frame of the fiducial
observer. We would need to fit these local Minkowski spaces
together and would then get the curved spacetime.

The normalized timelike vector field $n^\mu$, obeying $n^\mu n_\mu=-1$ and $n^\mu \gamma_{\mu\nu}=0$, is constructed in terms of the lapse function, $n_\mu=(\alpha,0,0,0)$ and $n^\mu=(-1/\alpha,0,0,0)$ (the shift vector is zero). Thus, the $3+1$ decomposition is achieved by projecting every tensor onto $n^\mu$ in timelike hypersurfaces and onto $\gamma_{\mu\nu}$ in spacelike hypersurfaces. For example , the metric is decomposed as $g_{\mu\nu}=\gamma_{\mu\nu}-n_\mu n_\nu$. We, now, proceed to decompose the relevant tensors in terms of  $n^\mu$ and $\gamma_{\mu\nu}$.

We first deal with the four-velocity $U^\mu=(\Gamma, \Gamma v^i)$ where $v^i=dx^i/dt$ corresponds to the $i$-component of the fluid velocity $\bm{v}$, and $\Gamma$ is the Lorentz factor.
Since $n_\mu U^\mu=\alpha \Gamma$, the decomposition
\begin{equation}
U^\mu=-\alpha \Gamma n^\mu+\Gamma{\gamma^\mu}_\nu v^\nu\, ,
\label{velocityU3+1}
\end{equation}
allows us to write the Lorentz factor as
\begin{equation}
\Gamma=\left(\alpha^2-\gamma_{\mu\nu}v^\mu v^\nu\right)^{-1/2}\, .
\label{gammacurved}
\end{equation}
In flat space using Cartesian coordinates $\alpha=1$,
$\gamma_{ij}=\delta_{ij}$, and the well-known Lorentz factor of special relativity
$\Gamma=(1-v^2)^{-1/2}$ is recovered.

The preceding definitions can be put in the local fiducial observer (FIDO) frame. For the FIDO, the plasma velocity
is given in terms of $\tau$, the FIDO proper time, as $v_F^i=dx^i/d\tau=\alpha^{-1}v^i$  \cite{thorne}. Thereby, the Lorentz factor measured by the FIDO is $\Gamma_F=\left(1-{v_F}^i {v_F}_i\right)^{-1/2}=\alpha \Gamma$. In this way, in the FIDO frame, $n_\mu U^\mu=\Gamma_F$. Though we will continue using the definition \eqref{gammacurved}, all our results can be put in the FIDO frame in a straightforward way.

With the nomenclature straightened out, and neglecting the plasma
back-reaction on spacetime, one may readily write down the decomposition of the field equations. To illustrate the procedure for subsequent calculations, we begin with the  Maxwell equations \eqref{Maxwellcurved}.
Several authors \cite{pla1,pla3,pla4,pla5,pla2,pla6,pla7,pla8,tajima,tarkenton,acht} have expressed the electromagnetic tensor in terms of the electric ($E^\mu$) and the magnetic ($B^\mu$) fields, defined as ($\epsilon^{\alpha\beta\gamma\delta}$ is the totally antisymmetric tensor)
\begin{equation}
 E^\mu=n_\nu F^{\nu\mu}\, ,\qquad B^\mu=\frac{1}{2}n_\rho \epsilon^{\rho\mu\sigma\tau}F_{\sigma\tau}\, .
\label{EBdecomposed}\end{equation}
Both fields are spacelike, $n_\mu E^\mu=0$ and $n_\mu B^\mu=0$, and allow the electromagnetic tensor
to be decomposed as
\begin{equation}
 F^{\mu\nu}=E^\mu n^\nu-E^\nu n^\mu-\epsilon^{\mu\nu\rho\sigma}B_\rho n_\sigma\, .
\label{FaradayD}
\end{equation}
Substituting \eqref{FaradayD} into \eqref{Maxwellcurved}, and projecting it onto $n_\mu$ we find ${E^\mu}_{;\mu}=4\pi q n \alpha\Gamma$, that translates into the scalar form as \cite{thorne,tajima,tarkenton}
\begin{equation}
 \nabla\cdot\bm E=4\pi q n \alpha\Gamma\, ,
\label{div1E}\end{equation}
where $\nabla$ is the spatial covariant derivative derived from $\gamma_{\mu\nu}$.
Projecting Eq.~\eqref{Maxwellcurved} onto ${\gamma^\beta}_\mu$, we find the spacelike equation ${\gamma^\beta}_\mu {E^\mu}_{;\nu}n^\nu-\epsilon^{\beta\nu\rho\sigma}(B_\rho n_\sigma)_{;\nu}=4\pi q n \Gamma v^\beta$ which, using  $n_{\mu;\nu}=-n_\nu \alpha_{,\mu}/\alpha$ \cite{tarkenton}, yields
\begin{equation}
\frac{1}{\alpha} \nabla\times\left(\alpha\bm B\right)=4\pi qn\Gamma\bm v+\frac{1}{\alpha}\frac{\partial\bm E}{\partial t}\, ,
\label{div2E}\end{equation}
the GR modified Maxwell law \cite{thorne,tajima,tarkenton}.

Note that, from the preceding two equations, we can derive the continuity equation
\begin{equation}
\frac{\partial}{\partial t}\left(\alpha n \Gamma\right)+\nabla\cdot\left(\alpha n\Gamma \bm v\right)=0\, ,
\end{equation}
that could, just as well, be obtained when the decomposition \eqref{velocityU3+1} is introduced in the covariant equation $(nU^\mu)_{;\mu}=0$.

For the homogeneous Maxwell equations, one defines the dual electromagnetic tensor
\begin{eqnarray}
 F^{*\mu\nu}=\frac{1}{2}\epsilon^{\mu\nu\rho\tau}F_{\rho\tau}=B^\mu n^\nu-B^\nu n^\mu+\epsilon^{\mu\nu\rho\tau}E_\rho n_\tau\, ,
\end{eqnarray}
that satisfies ${F^{*\mu\nu}}_{;\nu}=0$ by its antisymmetry. When projected onto $n_\mu$,
we find the timelike decomposition ${B^\nu}_{;\nu}=0$, alternatively  written as \cite{thorne,tajima,tarkenton}
\begin{equation}
 \nabla\cdot\bm B=0\, .
\label{div1B}
\end{equation}
The spacelike projection, ${\gamma^\beta}_\mu{B^\mu}_{;\nu}n^\nu+\epsilon^{\beta\nu\rho\tau}(E_\rho n_\tau)_{;\nu}=0$, has the vectorial equivalent \cite{thorne,tajima,tarkenton}
\begin{equation}
\frac{\partial\bm B}{\partial t}=-\nabla\times\left(\alpha\bm E\right)\, ,
\label{div2B}
\end{equation}
 the GR version of Faraday's law.
 Equations.~\eqref{div1E}, \eqref{div2E}, \eqref{div1B} and \eqref{div2B} constitute the Maxwell's equations in curved spacetime in the  $3+1$ decomposition. They are  rather similar to Maxwell's equations in flat space: the spacetime curvature effects enter through the lapse function $\alpha$.

Now we turn to the $3+1$ formulation of the total unified dynamics of the magnetofluid embedded in curved spaacetime.  Because of the antisymmetry of $M^{\mu\nu}$, the decomposition will be analogous  to that for $F^{\mu\nu}$. In terms of the generlized electric ($\xi^\mu$) and magnetic ($\Omega^\mu$) fields
\begin{equation}
 \xi^\mu=n_\nu M^{\nu\mu}\, ,\qquad \Omega^\mu=\frac{1}{2}n_\rho \epsilon^{\rho\mu\sigma\tau}M_{\sigma\tau}\, ,
\end{equation}
both spacelike ($n_\mu \xi^\mu=0$ and $n_\mu \Omega^\mu=0$), the magnetofluid tensor reads
\eqref{FaradayD},
\begin{equation}
 M^{\mu\nu}=\xi^\mu n^\nu-\xi^\nu n^\mu-\epsilon^{\mu\nu\rho\sigma}\Omega_\rho n_\sigma\, .
\label{Mdecomposicion}
\end{equation}

The detailed form of the vector fields $\xi^\mu$ and $\Omega^\mu$ can be worked out using the definition $M^{\mu\nu}=F^{\mu\nu}+(m/q)S^{\mu\nu}$, and the four-velocity \eqref{velocityU3+1}; the three
-vector  generalized electric and magnetic fields
\begin{equation}
 \bm\xi=\bm E-\frac{m}{\alpha q}\nabla\left(f\alpha^2\Gamma\right)-\frac{m}{\alpha q}\frac{\partial}{\partial t}\left(f\Gamma\bm v\right)\, ,
\label{GE}\end{equation}
\begin{equation}
\bm\Omega=\bm B+\frac{m}{q}\nabla\times\left(f\Gamma\bm v\right)\, ,
\label{GVO}\end{equation}
are the curved spacetime generalization of the corresponding vector fields defined in Refs.~\cite{mah1,mah12}. We remind the reader that, in our usage, the  generalized magnetic field $\bm \Omega$ is synonymous with the generalized vorticity, GV. Evidently, general relativity enters the definition of GV through $\Gamma$.

Substituting the fields ($\xi^\mu$ and $\Omega^\mu$) into \eqref{Mdecomposicion}, the covariant equation of motion \eqref{eqmo} converts to
\begin{equation}
 \alpha \Gamma \xi^\mu-\Gamma v_\nu \xi^\nu n^\mu-\Gamma v_\nu\epsilon^{\mu\nu\rho\sigma}\Omega_\rho n_\sigma=\frac{T}{q}\sigma^{,\mu}\, ,
\label{eqmoGR}
\end{equation}
from which the  $3+1$ equations are obtained by appropriate projections on the timelike and spacelike hypersurfaces. The  $n^\mu$ projection, $\Gamma v_\mu\xi^\mu=(T/q)n_\mu\sigma^{,\mu}$, is the equation for energy conservation
\begin{equation}
q \alpha \Gamma \bm v\cdot\bm \xi=-T\frac{\partial\sigma}{\partial t}\, ,
\label{energyCon}\end{equation}
while the  ${\gamma^\beta}_{\mu}$ projection, $\alpha\Gamma \xi^\beta+\Gamma n_\tau\epsilon^{\tau\beta\nu\rho}v_\nu\Omega_\rho=(T/q)\sigma^{,\beta}$, yields the momentum evolution equation
\begin{equation}
 \alpha\Gamma\bm\xi+\Gamma\bm v\times\bm\Omega=\frac{T}{q}\nabla\sigma\, .
\label{momentCon}
\end{equation}

The charge $q$ (and mass $m$),
referring to the constants attributes of the ``particles"  that make up the fluid,
pose no conceptual problems when we ignore the back reaction.  Thermodynamic
quantities like temperature, $T$, entropy density, $\sigma$, and
$f$ are more problematic. One would need to formulate more
clearly what they signify in the strong field region near, for instance, the
surface of a black hole. In the current work, we assume that thermodynamical properties belong to the "test matter" plasma, where the normal definitions are adequate
in the chosen frame.

Equations \eqref{energyCon} and \eqref{momentCon} may look somewhat unfamiliar. However, it is possible to show that they are equivalent to the usual $3+1$ plasma equations \cite{tajima,tarkenton} invoked in plasma literature. For example, the effects of  the interaction of the fluid with the local gravitational acceleration are hidden in the definition of the unified fields. This is spelled out in Appendix A.

There is a very strong reason for the use of Eqs.~\eqref{energyCon} and \eqref{momentCon} instead of other extant formalisms. The unified magnetofluid approach, epitomized in ~\eqref{energyCon} and \eqref{momentCon}, is a very powerful tool that leads us directly to the general vortical form of the plasma equations. It is in this form, that the sources of general vorticity (magnetic fields being a part) are explicitly revealed, and it becomes relatively easy to develop an encompassing theory for the  generation of general vorticity. Equations ~\eqref{energyCon} and \eqref{momentCon} are expected to be as effective in isolating the sources of  vorticity in curved spacetime  as their special relativistic antecedents.

We have  yet to derive  the promised ``vortical'' plasma system in curved spacetime. The antisymmetry of the unified tensor $M^{\mu\nu}$, in analogy with $F^{\mu\nu}$, implies that its dual must obey ${M^{*\mu\nu}}_{;\nu}=0$. The $3+1$ decomposition of this equation, equivalent to Eqs.~\eqref{div1B} and \eqref{div2B}, will lead to a spacelike projection
\begin{equation}
\frac{\partial\bm \Omega}{\partial t}=-\nabla\times\left(\alpha\bm \xi\right)\, .
\label{div2BM}
\end{equation}
When the constraint \eqref{div2BM} is used in Eq.~\eqref{momentCon}, we arrive at
\begin{equation}
\frac{\partial\bm \Omega}{\partial t}-\nabla\times\left(\bm v\times\bm\Omega\right)={\bf \Xi}_B+{\bf \Xi}_R\, ,
\label{seedmagnetic}
\end{equation}
that has, precisely, the standard vortical form. ${\bf \Xi}_B$ and ${\bf \Xi}_R$, explicitly displayed on the right hand side, are the possible sources of the vorticity $\bm\Omega$. Both these drives are nonzero only for inhomogeneous thermodynamics. The first one is the traditional baroclinic term  \cite{mah1,mah12}
\begin{equation}
{\bf \Xi}_B=-\left(\frac{1}{q \Gamma}\right)\nabla T\times\nabla
\sigma\, ,
\label{generclasico}
\end{equation}
corrected by curvature. The non-relativistic limit of this term, called the  Biermann battery, has been extensively studied. The second term, the general relativistic drive
\begin{eqnarray}
{\bf \Xi}_R&=&\frac{T}{q \Gamma^2}\nabla\Gamma\times\nabla\sigma \nonumber\\
&=&\frac{T\, \Gamma}{2q}\left[-\nabla \alpha^2+
\nabla\left(\gamma_{ij}v^iv^j\right)\right]
\times\nabla\sigma\, ,
\label{generB}
\end{eqnarray}
is the principal object of this search. The terms ${\bf \Xi}_B$
and ${\bf \Xi}_R$ are the non-magnetic thermodynamic source
terms that create the  conditions for driving the linear growth
of the magnetic fields from a zero initial value. In this
sense, these drives act as batteries.

Although the baroclinic term given in \eqref{generclasico} is
somewhat modified by the curved spacetime metric $g_{\mu\nu}$
(through $\Gamma$ and $\nabla$), there is no dramatic or qualitative change.
The relativistic drive ${\bf\Xi}_R$, however, is radically
transformed from its flat space antecedent \cite{mah1,mah12} to
which it reduces in the appropriate limit. The striking result
is that {\it the gravitational potential, through $g_{00}$}
({\it or $\alpha$}), {\it can produce a magnetic field in any region
populated by charged particles even if their local velocities
are negligible; it could be called a Gravito-Magnetic
 battery}!

We expect this result to have many astrophysical consequences. In particular, we can compare the strengths of the baroclinic term and the general relativistic drive. If the baroclinic drive is nonzero, then
\begin{equation}
  \frac{|{\bf\Xi}_R|}{|{\bf\Xi}_B|}\approx \frac{l\,  \Gamma^2}{\sqrt{1-r_0/r}}\left|\nabla\alpha^2-\nabla(\gamma_{ij}v^i v^j)\right|\, ,
\end{equation}
where $l/\sqrt{1-r_0/r}$ is the scale length of variation of the temperature corrected by the curvature. If $l$ is similar to the scale length of the variations of the relativistic effects of the plasma, then the general relativist drive can be much more important than the baroclinic term when $\Gamma^2\gg 1$ (i.e. when $\alpha^2-\gamma_{ij}v^i v^j\ll 1$). Note that the general relativistic drive can be relevant even if the plasma velocities are negligible in some special case configuration. In conclusion, when the plasma is under strong gravitational fields and/or high relativistic effects, the general relativistic drive is the more relevant source for the magnetic field generation. In flat spacetime, the relativistic drive will be important only for relativistic  velocities \cite{mah1}.

Before estimating the magnitude of the new drive, we would like to emphasize that this first conceptual paper will be limited to
demonstrate the existence of a
curvature-driven drive for vorticity/magnetic field. More detailed and rigorous calculations
and their consequences, will be submitted in a subsequent paper.
It is true that a simple-minded extension of  special relativistic notions
to curved spacetime can cause conceptual problems. There are two ingredients
required for the special relativistic mechanism to work: (a) an inhomogeneous
stress-energy tensor; and (b) a preferred direction provided by
the ``boost". When we go to, say, a Schwarzschild spacetime,
these ingredients are missing. The following prescription
provides the proper framework for extending the formalism to
curved spacetime. We can put the inhomogeneity into some non
self-gravitating ``test matter" that has been neglected
compared with the mass of the Schwarzschild entity. We must,
similarly, rely on the ``test matter" to provide the second
ingredient. Both would be provided, for example, by the plasma near the black hole.

\section{Estimates for vorticity generation}
\label{aplicacion}

Though the main result of this paper is the analytic expression \eqref{generB},
we will now estimate the strength of the relativistic drive and vorticity/magnetic field
in a very simplified model of a plasmas in curved space-time.
Consider an accretion plasma disk around a Schwarzschild black hole.
The relevant space-time metric elements are: $\alpha^2=1-r_0/r$,
$\gamma_{rr}=\alpha^{-2}$, $\gamma_{\theta\theta}=r^2$
and $\gamma_{\phi\phi}=r^2\sin^2\theta$, where $r_0=2MG/c^2$ is the Schwarzschild radius, $M$ is the mass of
the black hole, $G$ is the gravitational constant, and $r$ is
the radial distance to the plasma matter (from now on we will put $c$ explicitly in the calculations).
We will estimate the GV in two representative cases: 1) for a plasma element in a stable  orbit at $5r_0$, and 2) for a free-falling plasma element  near the Schwarzschild radius.

\subsection{Seed generation in an accretion plasma}

We assume that the plasma is in a thin accretion disk, and moves in the equatorial plane ($\theta=\pi/2$) of the disk with zero azimuthal speed, $\dot{\theta}=0$. For an in-spiral motion for thin disks, the orbital velocity can be estimated to be Keplerian,  $v^\phi=r\dot\phi=c \sqrt{r_0/2 r}$; we will assume it   to be larger than the radial velocity at which matter falls into the black hole, $v^\phi\gg v^r$ \cite{vietri,shapiro}. We can only ensure this sufficiently far from the gravitational
source and would, therefore, miss the really strong-field effects. To ensure relatively stable
orbits about the black hole, we will locate the plasma disk at about $5 r_0$, where we could neglect the
the radially inward component of the velocity.

In addition to the spatial variations of the
metric tensor, the drive ${\bf \Xi}_R$ depends on the gradients
of the entropy density.
At $5 r_0$, the usual definition for entropy \cite{mah2}
is valid since the nonlinearity of the gravitational field is not dominant,
If, in addition, the plasma obeys a barotropic equation of state, i.e, the pressure is
a function of density,  $\sigma=F(T)$,
then  $(T/c)\nabla\sigma\equiv \zeta k_B \nabla T$
where $\zeta$ is  of order unity.

Note that for a plasma with this kind of an equation of state, the baroclinic drive ${\bf
\Xi}_B$ vanishes because $\nabla\sigma\propto\nabla T$; the only source left for
generating a magnetic field is the general relativistic drive.

In the $3+1$ decomposition, the gradient of a scalar field $P$, $\nabla P=(1-r_0/r)^{1/2}\, \partial_rP\, \hat e_r+(1/r)\partial_\theta P\, \hat e_\theta+(1/r\sin\theta)\partial_\phi P\, \hat e_\phi$, has the the unusual factor $(1-r_0/r)^{1/2}$ coming from the radial metric coefficient.
For the model described above, the general relativistic drive
(in the equatorial plane $\theta=\pi/2$) becomes
\begin{equation}\label{generB20}
  {\bf\Xi}_R=\frac{3\zeta c k_B r_0 \alpha}{4\, e\, r^3}\left(1-\frac{3 r_0}{2 r}\right)^{-1/2}\frac{\partial T}{\partial\phi}\hat e_z\, ,
\end{equation}
where the variations of the temperature have been taken in cylindrical geometry, we have used the electron charge $q=-e$, and we have simplified the thin disk model by neglecting the toroidal
temperature gradients compared with the poloidal variations, $\partial_\theta T\ll\partial_\phi T$.

All the charged matter of the accretion disk contributes to
${\bf\Xi}_R$, and therefore acts as a source for ${\bm \Omega}$. Since
${\bf\Xi}_R\rightarrow 0$ for $r\rightarrow \infty$, the
contribution  from matter relatively close to the compact object will be dominant. Notice that the relativistic drive has a net flux in the $\hat e_z$ direction. For the stable orbit at $r=5r_0$, the relativistic drive \eqref{generB20} simplifies to
\begin{eqnarray}
{\bf \Xi}_R\approx\frac{3\zeta c k_B}{500\, e\, r_0^2}\frac{\partial T}{\partial\phi}\, \hat e_z\, ,
\label{generB2}
\end{eqnarray}
and is proportional to the temperature of the disk. We can assume that the complete accretion disk radiates like a blackbody with an average temperature $\bar T= \int \partial_\phi T d\phi\approx 5\times 10^7 (M_\odot/M)^{1/4}$K \cite{vietri}, where $M_\odot$ is the solar mass. It is easy to see that as long as the black hole mass $M\geq 10^{-2} M_\odot$, $x=mc^2/k_B T\gg 1$, and the plasma temperature remains non-relativistic.

Under these conditions, the total relativistic drive \eqref{generB2} produced by the plasma in the thin ring
of matter centered around $r=5r_0$, can be estimated as
\begin{eqnarray}
{{\bf \Xi}_R}_{\begin{scriptsize}\mbox{total} \end{scriptsize}}
&=&\int_0^{2\pi}d\phi\, \, {\bf\Xi}_R\nonumber\\
&\approx&3\times 10^{-2}\zeta\left(\frac{M_\odot}{M}\right)^{9/4}\hat e_z\, .
\end{eqnarray}

Substituting the simplified drive into
Eq.~\eqref{seedmagnetic}, the GV generated by the space-time
curvature can be calculated. Let us begin with an initial
state with zero GV.  For some short enough time $\varsigma$
(the initial seed generation phase), when the nonlinear terms
involving ${\bm \Omega}$ are negligible, ${\bm \Omega}$ grows
linearly with time: ${\bm \Omega}_{\begin{scriptsize}\mbox{total} \end{scriptsize}}\approx {{\bf
\Xi}_R}_{\begin{scriptsize}\mbox{total}
\end{scriptsize}}\varsigma$.
To estimate the growing time $\varsigma$, we notice that the linear proportionality
cannot hold when the nonlinear term in \eqref{seedmagnetic} is
comparable to ${{\bf \Xi}_R}_{\begin{scriptsize}\mbox{total}
\end{scriptsize}}$. A good measure of  $\varsigma$  is provided by
the relation $|{\bm \Omega}_{\begin{scriptsize}\mbox{total}
\end{scriptsize}}|\varsigma^{-1}\simeq |\nabla\times({\bm v}\times{\bm \Omega}_{\begin{scriptsize}\mbox{total}
\end{scriptsize}})|$ implying that $\varsigma\simeq
 L/|{\bm v}|$, where $L$ is the length of
variation of the $|{\bm v}\times{\bm \Omega}|$ force.
Taking the  length $L$ on which $|{\bm v}|$ varies to be of the order of the (curvature corrected) variation scale $5r_0/\alpha$, the time for initial linear phase of GV
seed formation may be approximated as
\begin{eqnarray}
\varsigma=\frac{5r_0}{|\bm v|\alpha}\approx 1.7\times 10^{-4}\left(\frac{M}{M_\odot}\right)\, ,
\end{eqnarray}
measured in seconds, where we have assumed that the velocity is of order $v^\phi$. Thus, the total strength of the magnetic field generated (in gauss) for the ``test" plasma matter accreting at a distance $5r_0$ is
\begin{equation}\label{MagTotalfinal}
| {\bf \Omega}_{\begin{scriptsize}
\mbox{total}\end{scriptsize}}|\approx{5\times 10^{-6}}\zeta\left(\frac{M_\odot}
{M}\right)^{5/4}\, ,
\end{equation}
and lies in the $\hat e_z$ direction. For a black hole of stellar mass ($M\approx M_\odot$), the maximum generated  magnetic field seed is found to be of the order of $|{\bf \Omega}_{\begin{scriptsize} \mbox{total}\end{scriptsize}}|\approx 5\times10^{-6}$G.

It is important to realize that this initial seed is supposed to be small. It is what is created in a very short initial time in a state where there was, precisely, no magnetic field to begin with. The existence of this seed  is crucial to the very startup of the standard processes of long-time magnetic field generation, like the dynamo process or the magneto-rotational instability. The dynamo process that converts short scale fluid vorticity into long term magnetic field (electromagnetic vorticity) can operate only when it has some initial magnetic field to amplify; we have just shown that the General Relativistic drive can, precisely, provide the needful.

\subsection{Strong field generation near the horizon}

Using the appropriate simplified version of
formula \eqref{MagTotalfinal}, we just estimated the small seed magnetic field that the GR drive can generate in an accretion disk around a black hole black. One naturally expects that the GR drive will get considerably stronger as our test plasma moves closer and closer to the event horizon at $r_0$. To get an idea of the strength of the drive, here we do a very simple, somewhat crude, calculation. A more sophisticated treatment, including various QED plasmas effects \cite{chou}, is left for future work.

When the plasma is near the horizon (there are no stable particle orbits), we may approximate it as a fluid in free fall with a purely radial velocity $v^r$. As the fluid element approaches the horizon (in $r=r_0$), the radial velocity and the Lorentz factor  roughly go, respectively,  as $v^r\approx c\alpha^2\sqrt{r_0/r}$ (measured in the universal time) and $\Gamma\approx1/\alpha^2$ , so that $\Gamma v^r\approx c \sqrt{r_0/r}$ \cite{shapiro}. Then, the relativistic drive \eqref{generB} is
\begin{equation}\label{}
  {\bf \Xi}_R\approx\frac{\alpha \zeta c k_B}{e r_0^2}\frac{\partial T}{\partial\phi}\hat e_z\, .
\end{equation}
Notice that the drive, as always, is  inhomogeneity-driven and needs a non-radial gradient of the plasma temperature.  The growing time $\varsigma$ may be estimated like we did  in the farther accretion region. As the plasma location approaches the horizon, the growing time $\varsigma\approx r_0/(c\alpha^3)$, leading to a simple estimate
\begin{equation}\label{GVhorizon}
|{\bf \Omega}_{\begin{scriptsize} \mbox{total}\end{scriptsize}}|\approx\frac{\zeta k_B T}{\alpha^2 e r_0 }\, ,
\end{equation}
for the total GV generated. Despite the crude approximations invoked to obtain the total GV \eqref{GVhorizon}, we have found quite a spectacular result. Since $\alpha^2=1-r_0/r \rightarrow 0$ near the horizon, enormous  GV (magnetic field) can be generated by the mechanism investigated in this paper. This mechanism, if it survives more thorough examination (via, perhaps, detailed numerical calculations), could  provide just the strong guide field that could collimate escaping plasma particles and advance our understanding of the formation of astrophysical jets. The temperature $T$ in \eqref{GVhorizon} is only a perturbation of the homogenous spherical symmetric temperature of the free-falling plasma \cite{shapiro}. We assume that this kind of perturbations will always be present. Even  for small  temperature perturbations, the result \eqref{GVhorizon} shows that the GV can be very large near the horizon.

\section{Discussion}
\label{discusion}

We have demonstrated the existence of a new Gravito-Magnetic battery
mechanism [with strength given by a source term \eqref{generB}]
for generating the seed vortex/magnetic field in astrophysical and cosmic settings.
The battery action is created by a fundamental interaction of gravity (causing space
time curvature) and inhomogeneous plasma thermodynamics; both
elements are essential.
The current theory is quite unlike other classical
theories that invoke, for instance, the difference in the $e/m$
ratio between protons and electrons to create initial currents
or those that introduce drag effects in the electron motion (Compton drag)
to create initial currents in the context of cosmology
\cite{vietri}.
Besides, the Gravito-Magnetic battery mechanism presented here has the advantage that it is nonzero when the standard Biermann battery is null. The Biermann battery, driven by the baroclinic term, is rather difficult to operate because the variations of temperature and entropy tends to align in thermodynamical equilibrium, $\nabla T\times\nabla\sigma=0$. It is likely that in most cases of interest, the general relativistic drive would be the only source to produce seed vorticity and magnetic fields.

Though the most important result of this paper is contained in the analytic forrmula Eq.~\eqref{generB}, we have chosen to explicitly estimate the strength of the generated vorticiy for two representative cases ; 1) The plasma is an accretion disk located around $5r_0$ from the black hole. In this relatively weak field region with stable particle orbits, one is interested in calculating the seed field that could be a progenitor, for instance, of a dynamo action, 2) the plasma is in free fall near the horizon; the idea is to see if a strong enough magnetic field can be created for jet formation.

For the first scenario, two explanatory remarks are in order: 1) Parity breaking in the
gravitational field, involved in generating a magnetic field,
need not be  worrisome because of the opposite parities of the
gravitational and electromagnetic fields; 2) there is no
guarantee that the  particles in the stable orbit
(at $r=5 r_0$) will not fall into the black hole. We follow
here the standard assumption made in Astrophysics that though
some matter will escape from the stable region, other matter
will replace it. In that sense one could think of invoking
above estimates for orbits closer than this limit. However, the timescale for infall, and
the breakdown of the assumption that the speed of infall is
negligible, prevents such an extension.

It is also worth while to contrast our mechanism with
relativistic effects like the Blandford-Znajek (B-Z) \cite{bz},
used in modeling active galactic nuclei, quasars and gamma ray
bursters; the latter deals with strong fields, highly energetic
events. By contrast, this seed creating mechanism pertains  to a
test plasma in a relatively stable orbit around the black hole. Further,
in our analysis it is the plasma that is ``rotating" while in the B-Z case it is the
gravitational source that is spinning.

The vorticity/magnetic field obtained in Eq.~\eqref{MagTotalfinal} is rather small.
However it is more than adequate as a crucial seed field to drive
a dynamo amplification. Gravity, in this, case just gets the process started, the
eventual energy for field generation in the accretion disk comes from short scale velocity turbulence.

The literature is full of mechanisms, explored only for the purpose of
producing small seeds of magnetic fields: the rotation of black holes \cite{khanna,leahy}, and the  radiation force on electrons \cite{kogan}, being two examples.
As we said before, during the short phase where nonlinear
effects can be neglected, the magnetic field can grow (from a state of zero field) linearly
with time. Once created, these seeds can grow further by a variety of
nonlinear processes. The dynamo mechanism is, of course, one of
the most investigated mechanisms for black holes \cite{pudritz,colgate}, where the rotation of the black hole can introduce a new effect which is added to the known $\alpha-\Omega$ dynamo \cite{khanna2,rein}.
Nonlinear effects can, in addition,
provide long range order to the generated magnetic fields. Well
known examples are: the {\it shearing} of the magnetic field,
and Parker's mechanism \cite{parker}. Both these ideas pertain
to rotating objects in which poloidal (toroidal) magnetic field
lines transform into toroidal (poloidal) ones.

The motivation for the second part of the calculation, where we deal with a test plasma in the vicinity of the horizon, is entirely different. A  very rough estimate shows that the Gravito-Magnetic drive \eqref{GVhorizon} turns out to be  very strong  in this neighborhood. A detailed study, however, is needed: 1) to calculate the long time evolution of the growth of GV because the initial stage of linear growth will soon yield to the nonlinear stage, and 2) to incorporate other effects  than those considered in this work \cite{chou}. Based on our rough estimates, we can certainly argue  that the general relativistic drive can be  a source of large magnetic fields. Here, unlike the accretion disk case, it is  gravity that is directly feeding GV and the magnetic field. The curvature driven magnetic field (very near the horizon) may be just what we need for collimating jets of charged matter emitted from the accretion disk of compact objects. Again there are a variety of mechan!
 isms proposed to explain jet collimation; these mechanisms make varied
assumptions about the plasmas, the compact objects or the inertial effects of
the jet \cite{vietri,koideS}. Discussing the jet collimation within the framework of this model will be taken up  in future work.

The final magnetic field [the general
relativistic drive \eqref{generB}] is created intrinsically by
the curvature in combination with the properties of the plasma
matter accreting onto the black hole. However, we have not, yet,
examined the possibility of a black hole acquiring a magnetic
field due to the magnetized matter falling into it (see the
membrane paradigm \cite{thorne}). Even more exciting is the
possibility of the magnetic field being generated without an
accretion disk plasma. There is reason to believe that
astrophysical black holes spin \cite{aan,yk,gl}. A spinning black
hole would provide the preferred direction and cause neutral
matter to get ionized and become a magnetic plasma. The
procedure adopted here, of using the rest frame, would no
longer be available due to frame-dragging \cite{misner}.
The generalization of the previous $3+1$ decomposition to the Kerr metric will be used in that case. We find that the rotation of the black hole contributes to the general relativistic drive. Similar results for magnetohydrodynamics have been suggested~\cite{khanna}.
On the other hand, it would not be necessary to use the full Kerr metric
for a slowly rotating black hole. We could use the
Lense-Thirring effect \cite{lt,ci1,ci2} for a semi-classical analysis
without too much additional complication.
This, too,  is left for future work.

\begin{acknowledgments}

The work of SMM was supported by USDOE Contract No.DE-- FG 03-96ER-54366. FAA thanks the CONICyT-Chile for his Becas Chile Postdoctoral Fellowship.

\end{acknowledgments}

\appendix
\section{}

We can write the energy conservation equation~\eqref{energyCon} and the momentum equation \eqref{momentCon} in terms of fluid variables instead of unified fields. Using the definition \eqref{GE} for $\bm\xi$, the energy conservation equation becomes
\begin{equation}
\frac{1}{\alpha}\frac{\partial e}{\partial t} +\frac{1}{\alpha}\nabla\cdot\left(h\alpha^2  \Gamma^2\bm v\right)=qn\Gamma\bm E\cdot\bm v-h\Gamma^2\bm v\cdot\nabla\alpha\, ,
\end{equation}
where the energy density is $e=h\alpha^2\Gamma^2-p$. Notice that the last term is the interaction of the fluid with the local gravitational acceleration.

In the same way, using \eqref{GVO} for $\bm\Omega$, the momentum equation could be written as
\begin{eqnarray}
&& \frac{1}{\alpha}\left(\frac{\partial}{\partial t}+\bm v\cdot\nabla\right)\left(\alpha h \Gamma^2\bm v\right)=qn\alpha\Gamma\bm E+qn\Gamma\bm v\times\bm B\nonumber\\
&&\qquad\qquad\qquad-\frac{\nabla\left(\alpha p\right)}{\alpha}-h\Gamma^2 \bm v \left(\nabla\cdot\bm v\right)-\frac{e\nabla\alpha}{\alpha}\, .
\end{eqnarray}
This equation resembles the form of the plasma fluid dynamical equation in special relativity, where now the effects of general relativity are introduced via the lapse function and the $\Gamma$ factor. Again, the gravitational acceleration effect is in the last term.

The preceding two equations can as well be obtained using the formalism developed in Refs.~\cite{tajima,tarkenton}. In this case, the starting point is the $3+1$ decomposition of the plasma energy-momentum tensor
\begin{equation}
 T^{\mu\nu}=en^\mu n^\nu-n^\mu s^\nu-n^\nu s^\mu+W^{\mu\nu}\, ,
\label{enemomenape}\end{equation}
where $s^\mu=\alpha\Gamma^2{\gamma^\mu}_\nu v^\nu$ is the energy flux, and $W^{\mu\nu}=h\Gamma^2{\gamma^\mu}_\beta{\gamma^\nu}_\phi v^\beta v^\phi+p\gamma^{\mu\nu}$ is the stress tensor.
This energy-momentum tensor is also obtained when the decomposition for the four-velocity \eqref{velocityU3+1} is used in \eqref{GRenmomentensor}.

\end{document}